%% file: ibrcomp.tex
\documentclass[11pt]{article}
                       
\usepackage[margin=1.25in]{geometry}
\usepackage{times}
\usepackage{natbib}
\usepackage{amsmath}   
\usepackage{amssymb}  
\usepackage{amsthm}  
\usepackage{authblk}  
\usepackage{rotating}
\usepackage{subfigure}
\usepackage{algorithm}
\usepackage{algpseudocode}

\newcommand{\SC}{\ensuremath{\mathcal{SC}}}
\newcommand{\SN}{\ensuremath{\mathcal{SN}}}
\newcommand{\strats}{\ensuremath{S}}
\newcommand{\stratcand}{\ensuremath{\mathbf{C}}}
\newcommand{\solcand}{\ensuremath{\mathbf{X}}} 
\newcommand{\soln}{\ensuremath{x}} 
\newcommand{\policy}{\ensuremath{\pi}}
\newcommand{\pparam}{\ensuremath{\pi}}
\newcommand{\pdomain}{\ensuremath{\Pi}}
\newcommand{\rthresh}{\ensuremath{\tau}} 
\newcommand{\minsamp}{\ensuremath{\nu}} 
\newcommand{\mincsamp}{\ensuremath{\nu^\star}} 
\newcommand{\compliance}{\ensuremath{\mathit{compliance}}} 
\newcommand{\Support}{\ensuremath{\mathit{support}}} 
\newcommand{\LocalSearch}{\ensuremath{\mathit{LocalVariations}}} 

\newcommand{\numnode}[1]{n_{#1}}
\newcommand{\numplayer}[1]{N_{#1}}
\newcommand{\cperisp}{\chi}

\providecommand{\abs}[1]{\lvert#1\rvert}
\newcommand{\agentfont}{\textsf}
\newcommand{\agnt}[1]{{\agentfont{#1}}}

\title{Analyzing Incentives for Protocol Compliance in Complex Domains: A Case Study of Introduction-Based Routing\thanks{Presented at the \emph{Twelfth Workshop on the Economics of Information Security}, Washington, DC, June 2013.}
}
\author{Michael P. Wellman, Tae Hyung Kim, and Quang Duong}
\affil{Computer Science \& Engineering, University of Michigan}
                                          
\begin{document}
\maketitle

\begin{abstract}  
Formal analyses of incentives for compliance with network protocols often appeal to game-theoretic models and concepts.
Applications of game-theoretic analysis to network security have generally been limited to highly stylized models, where simplified environments enable tractable study of key strategic variables.
We propose a simulation-based approach to game-theoretic analysis of protocol compliance, for scenarios with large populations of agents and large policy spaces.
We define a general procedure for systematically exploring a structured policy space, directed expressly to resolve the qualitative classification of equilibrium behavior as compliant or non-compliant.
The techniques are illustrated and exercised through an extensive case study analyzing compliance incentives for introduction-based routing.
We find that the benefits of complying with the protocol are particularly strong for nodes subject to attack, and the overall compliance level achieved in equilibrium, while not universal, is sufficient to support the desired security goals of the protocol.
\end{abstract} 

\section{Incentive Analysis of Network Protocol Compliance}   

Designers of a network security protocol typically start from an interest in properties satisfied by a system assuming all participants comply with the protocol.  This is often followed by consideration of whether compliance can be imposed or enforced in some way, or more flexibly whether autonomous agents would voluntarily comply, or whether they can be incentivized to do so.  
The behavior of interacting autonomous agents is the province of \emph{game theory}, 
which provides a general mathematical framework for characterizing rational decisions in a multiagent environment \citep{Leyton-Brown08}.  
Solution concepts such as Nash equilibrium offer a basis for predicting what rational agents will do when faced with strategic decisions.%
\footnote{This is not to say that such predictions will be very accurate, nor even the best one can do.
We consider equilibrium as a starting point, but advocate bringing in behavioral theories \citep{Camerer03} or indeed any other information that may be predictive for a particular scenario.
To a large extent we cast our methods in terms of general \emph{solution concepts}, so that they can be adapted to account for such information when it is available and formalizable in such terms.}    

Consider a system of \( N \) agents, where agent \( i \) chooses a strategy \( s_i \) from a set of available strategies \( \strats \).
(We do not need to assume symmetry, but do so now for descriptive simplicity.)  
A protocol can be defined as a partition of \( \strats \) into the \emph{compliant} strategies, \SC, and the \emph{non-compliant} strategies, \SN.%
\footnote{Our search procedure generalizes this to consider compliance a matter of degree, but we adopt the binary distinction here to simplify the motivating discussion.}  
A strategy profile \( s = (s_1,\dotsc,s_N) \) is compliant if \( s_i \) is in \SC\ for all \( i \).  
More generally, we may characterize the level of compliance of a (mixed) profile based on how many (in expectation) agents' strategies are compliant.
The straightforward way to address the compliance question for a particular environment is to solve the associated game with respect to some appropriate solution concept (e.g., equilibrium definition), and assess the (level of) compliance of solution profiles.

Game-theoretic modeling has seen increasing application to problems in networking \citep{Han11,Srivastava05} and security in particular \citep{grossklags08,Roy10,Vratonjic10}.
Some of this work addresses protocol compliance specifically, for example, \citet{Cagalj05} analyze incentives to comply with random packet deferment in CSMA/CA protocols.
They conclude that selfish nodes invariably violate the protocol in equilibrium, and propose methods to recover efficiency (essentially another level of protocol, for which compliance is an equilibrium) through enforceable cooperation among the violators.
In this and other studies, the game model is necessarily a highly stylized version of the underlying network scenario.
Well-crafted stylized models can produce valid and valuable insights, but always raise questions about whether their implications hold for richer environments that would be intractable for game-theoretic analysis.
Often, researchers further evaluate their results through simulation studies.
For example, \citet{Vratonjic12} model a content-monetization game as a two-player interaction between a web publisher and user, and then evaluate the derived publisher strategy in simulation over a heterogeneous population of users. 
Combining game-theoretic analysis with simulation can especially increase confidence when the simulation environment relaxes assumptions imposed for tractability in the game-theoretic model, as in the analysis by \citet{Chen07l} of wireless networks where nodes may tamper with contention windows.

For many network security protocols, it is not apparent how to define stylized environments that faithfully capture incentive issues of interest yet support tractable game-theoretic treatments.
For example, the set \( \strats \) of possible strategies may be too large (e.g., highly dimensional), or the scenarios as most naturally defined may include too many players, or excessively complex dynamics and information structures.
Hope always remains for clever modeling or algorithmic ideas to get beyond such obstacles, but in the meantime it is inappropriate to avoid systematic analysis of incentives in these domains.
We therefore pursue a simulation approach, not as mere adjunct of game-theoretic analysis, but as a \emph{means} to game-theoretic modeling.  
The basic idea is to explore a restricted space of heuristic strategies, employing simulation to estimate payoffs that can then be incorporated in a game model.

Because the search is (purposely) restricted, this technique should not be expected to precisely characterize solutions to the underlying game.
For the question at hand---protocol compliance---however, we are not necessarily interested in characterizing \emph{exactly} what the participating agents will do.
Rather, we ask the coarser question of whether their behavior is consistent with the specified protocol.                                           
The novel methodological contribution of this work is a procedure that directs exploration of strategy space toward answering this compliance incentive question.
                                    
The application driving our development of the procedure is a network protocol, \emph{introduction-based routing}, designed to incentivize responsible management of connection paths \citep{Frazier11}.
We briefly describe this protocol in the next section.
Section~\ref{sec:methods} follows with a detailed description of our empirical game-theoretic methods for analyzing protocol compliance in complex domains.
The results of applying these methods to the case-study domain are presented in Section~\ref{sec:ibr-analysis}.
Our study provides evidence for substantial but not universal compliance with the protocol in equilibrium.
The analysis further suggests that whereas the realized compliance level is sufficient to support the basic security goals of the protocol in these scenarios, additional incentives applied to nodes not directly subject to attack could improve overall compliance significantly. 

\section{Case Study: Introduction-Based Routing}

The \emph{introduction-based routing} (IBR) protocol represents a promising approach to deterring malicious behavior on networks \citep{Frazier11}.
By requiring an explicit connection before transmitting a message, IBR participants exert enhanced control over network activity.                      
New connections are formed exclusively through introduction by parties with existing connections to the requested endpoints, ensuring that there is a trail of responsibility behind every decision to accept a new communication partner.
Ensuring quality connections requires that participants pay attention to evidence of misbehavior, maintain reputation assessments for their connections, and propagate misbehavior reports throughout the network.
The protocol does not employ any form of global reputation mechanism, but rather relies on each node to maintain summary assessments of the nodes it has received information about.

The goal of strategic analysis in this domain is to establish whether network participants have adequate incentive to follow the rules of the IBR protocol. 
Even if one accepts the security claims of IBR proponents, the question of individual benefits to IBR compliance is an open question. 
An incentive analysis aims to determine whether or under what conditions nodes would indeed choose to comply, given full autonomy over that decision in service of their own interests.
\citet{AlK2013} show, under certain assumptions on attacks and detection, that the optimal policy for deciding whether to close or continue a connection based on feedback reports takes the form of a reputation threshold policy, just as IBR dictates.
This bolsters confidence in compliance plausibility, but leaves open several modes by which individual nodes may still choose not to comply, for example by failing to propagate attack information or indiscriminately introducing connections despite previous experience. 
Further modeling and analysis may illuminate the incentive tradeoffs for each element of the IBR policy, though this will likely prove tractable only for highly simplified network environments.

As established through extensive simulation studies \citep{Frazier11}, IBR networks are highly resistant to attack when all non-malicious nodes follow the protocol. 
We seek to validate in these same rich simulation environments that following the protocol is a reasonable behavior to expect from these nodes.
To do so, we first need to define what it means to comply with the protocol.
We start with the compliant policy recommended by the protocol designers, which includes a set of user-modifiable parameters, such as reputation thresholds for making and accepting introductions, thresholds for terminating existing connections, and amounts to increment or decrement reputations based on various forms of positive or negative feedback.  
Strategies in a specified region of this parameter space are classified as compliant (members of \SC), and those outside the region are considered non-compliant (\SN).

Whether IBR compliance is individually beneficial clearly depends on the context of compliance decisions made by others.
To reason about whether the IBR protocol is likely to be adopted by a community, the benefits at any given level of compliance by others can help us assess the plausibility of alternate adoption paths.
The game modeling approach pursued here can in principle provide a basis for reasoning not only about the existence of compliant equilibria, but also the dynamics of adoption that might lead to such equilibria.
                           
\section{Methods} 
\label{sec:methods}

\subsection{Approach: Empirical Game-Theoretic Analysis}   

In \emph{empirical game-theoretic analysis} (EGTA) \citep{Wellman06}, techniques from simulation, search, and statistics combine with game-theoretic concepts to characterize strategic properties of a domain.
The approach is a hybrid of analytical and agent-based modeling methods, and often appeals to evolutionary search techniques as well \citep{Phelps10}.
The basic EGTA step is simulation of a strategy profile, determining a payoff observation (i.e., a sample drawn from the outcome distribution generated by the simulation environment), which gets added to the database of payoffs.
Based on the accumulated data, we induce an empirical game model, typically (and in this work) a normal-form representation with payoffs estimated as the sample mean observed in simulation, for profiles that have been so evaluated.

The EGTA proceeds iteratively, with results from analyzing intermediate game models employed to guide the selection of further profiles to sample.
This naturally supports a dynamic approach to game formulation.
Whereas the full strategy space allowed by the simulator may be large or infinite, computational constraints limit the profiles for which we can obtain direct outcome observations.
Therefore, it makes sense to start from the most salient strategy candidates at first, incrementally adding candidates based on intermediate analysis results.
That is, we first solve a fairly restricted version of the game, admitting only a small slice of conceivable strategies. 
Based on these results, we then generate additional strategy proposals to be added to the candidate set. 
Further simulation and analysis produces solutions for an expanded game, which then represents the starting point for subsequent rounds of refinement.

\subsection{Compliance Search}

\begin{figure}[ht!]
	\centering
	\includegraphics[width=0.7\textwidth]{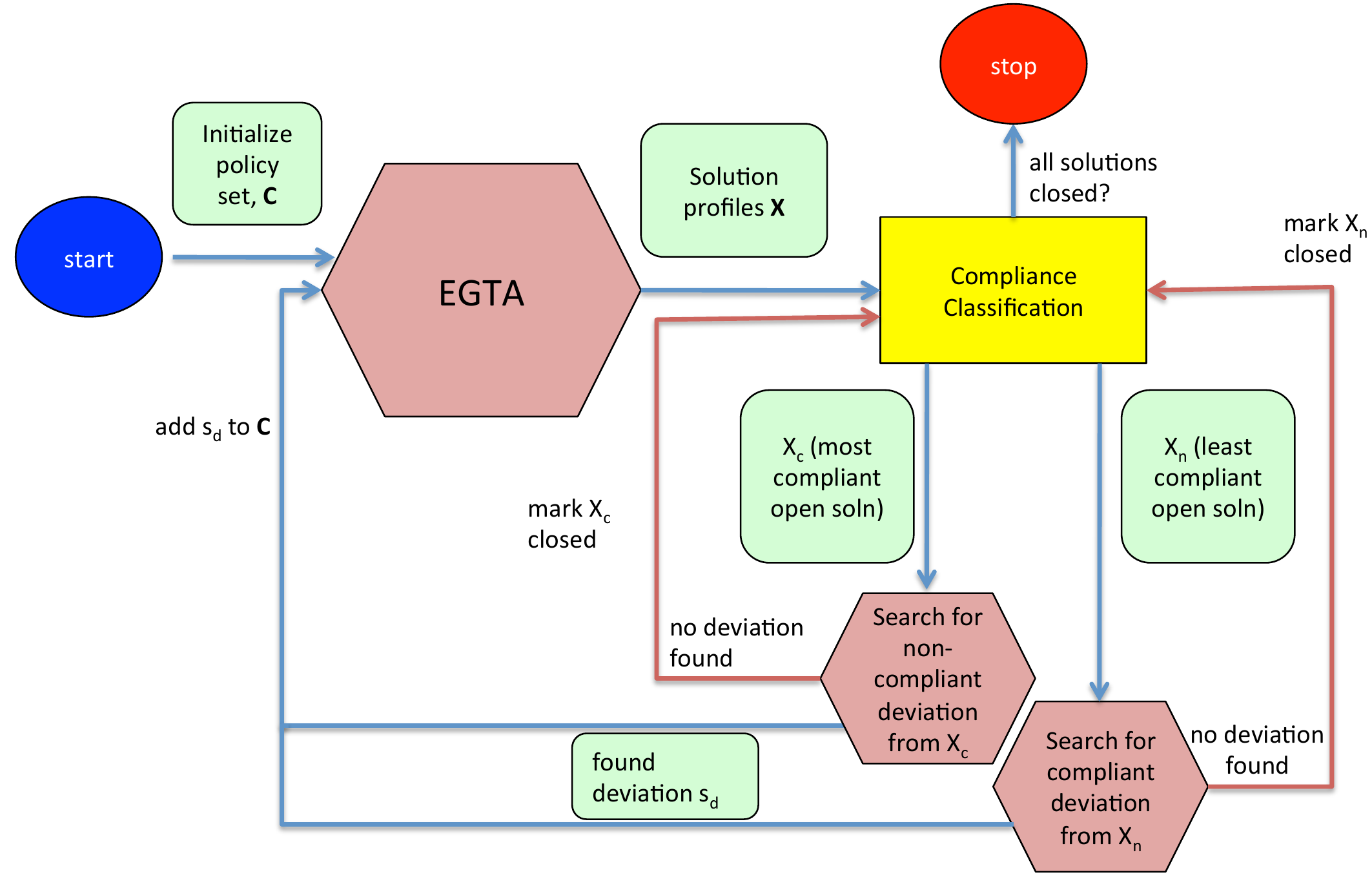}
	\caption{Compliance incentive analysis, searching over a large profile space guided by empirical game-theoretic analysis.}
	\label{fig:c-search}
\end{figure}

As indicated above, the main novelty in the current work is to focus EGTA on the goal of evaluating a particular characteristic of strategies---in this instance that of compliance with a network protocol.
The overall flow of the compliance search process is diagrammed in Figure~\ref{fig:c-search}.
The subprocesses represented by hexagons are detailed in Sections \ref{sec:inner} and~\ref{sec:outer}.

We start with a seed set of candidate strategies, \( \stratcand\subset\strats \).  
Our compliance incentive analysis proceeds iteratively, repeating the following steps: 
\begin{enumerate}
	\item Use EGTA (Figure~\ref{fig:inner-loop}) to derive a set of solution candidates (e.g., Nash equilibrium profiles) \solcand, using strategies only from \( \stratcand \).
	Elements of \solcand\ are considered \emph{open} until explicitly \emph{closed}.  
	\item  Let \( \solcand_c \) be the most compliant open solution in \solcand, and \( \solcand_n \) the most non-compliant (equivalently, least compliant).  
	\begin{enumerate}
		\item  Search \( \SN \setminus \stratcand \) for a beneficial deviation from \( \solcand_c \), using an appropriate strategy search method.  
		If no deviation can be found, mark \( \solcand_c \) as closed.
		\item  Search \( \SC \setminus \stratcand \) for a beneficial deviation from \( \solcand_n \), using an appropriate strategy search method.  
		If no deviation can be found, mark \( \solcand_n \) as closed.
	\end{enumerate}  
	\item  Add the beneficial deviation strategies \( s_d \) found in the previous steps to \stratcand.
\end{enumerate}
The procedure terminates when either all solutions in \solcand\ are closed, or (usually) we run out of time.  
The output is the set of solutions found, along with their open or closed status. 
We interpret the degree of compliance in output solutions as evidence for tendency to comply with the protocol.

The high-level version outlined above glosses a few details relevant to our implementation.
First, we assume that the agents are grouped into \( 1 \le R \le N \) roles, and that within each role the game is symmetric.%
\footnote{This \emph{role symmetry} assumption is without loss of generality, as we could always have a distinct role for each player (\( R=N \)).
The role construct allows us to exploit whatever player symmetry exists, up to full symmetry (all players in one role, \( R=1 \)).
As described below, our IBR analysis classifies network node players into \( R=4 \) roles.
}
Thus, in place of a single set of strategies \( S \) and strategy candidates, \stratcand, there are actually distinct sets of strategies \( S_r \) and strategy candidates \( \stratcand_r \) for each \emph{role}, \( r\in\{1,\dotsc,R\} \). 
The search for deviations is organized by role.
Second, whereas Figure~\ref{fig:c-search} shows the search for compliant and non-compliant deviations in parallel, we actually alternate between these two modes.

\subsection{EGTA Algorithm (Inner Loop)}
\label{sec:inner}  

The function of the EGTA inner loop is to identify and confirm one or more solution candidates, with respect to a fixed strategy space. 
We refer to this procedure as an \emph{inner loop}, as it operates iteratively as a subprocess (hexagon labeled ``EGTA'' in Figure~\ref{fig:c-search}) within the overall compliance search algorithm.
The general problem of selectively simulating a profile space for game-theoretic evaluation has been investigated by several researchers \citep{Fearnley13,Jordan08vw,Sureka:2005uq,Walsh03}.
Here we adopt a straightforward method that emphasizes systematicity of the search rather than optimization of simulation effort.

The solution concept we adopt is \emph{role-symmetric Nash equilibrium} (RSNE).%
\footnote{One could replace this with an alternative solution concept (based on game-theoretic equilibrium or not), with little or no change to the rest of the search process.
The completeness analysis imposes only the relatively weak assumption that a solution exists, so that an exhaustive exploration would eventually find it.
We also exploit the property that a solution \soln\ of the game is also a solution to any subgame for which \soln\ is defined.
This is true for RSNE (and NE-based concepts generally) in normal-form games, but solution concepts for which this fails may not be a good match for our procedure.
} 
A strategy profile is role-symmetric if, for all roles, each player in that role has the same (mixed) strategy. 
The \emph{support} of a mixed profile is the set of pure profiles played with positive probability. 
A \emph{deviation} of a profile \( q \) is a profile \( q' \) where all but one player plays according to \( q \).
A profile \( q \) is a Nash equilibrium if there exists no beneficial deviation: a deviation \( q' \) where the deviating player has greater expected payoff in \( q' \) than does its corresponding player in \( q \).
Every finite role-symmetric game has at least one RSNE.

Often we may be interested in approximate solutions.
An \( \epsilon \)-NE is a profile with \emph{regret} bounded by \( \epsilon \), where regret is the maximum gain to any player from deviating from the profile.
An exact NE has zero regret.
In general, for \( \epsilon > \epsilon' \ge 0 \), we can expect to be able to identify \( \epsilon \)-NE with less effort than required to identify \( \epsilon' \)-NE\@. 
Moreover, the regret of a profile provides a measure of its game-theoretic stability, which can be taken as one indicator of its plausibility as the choice of rational agents.

\begin{figure}[ht!]
  \centering
 	\includegraphics[width=0.65\textwidth]{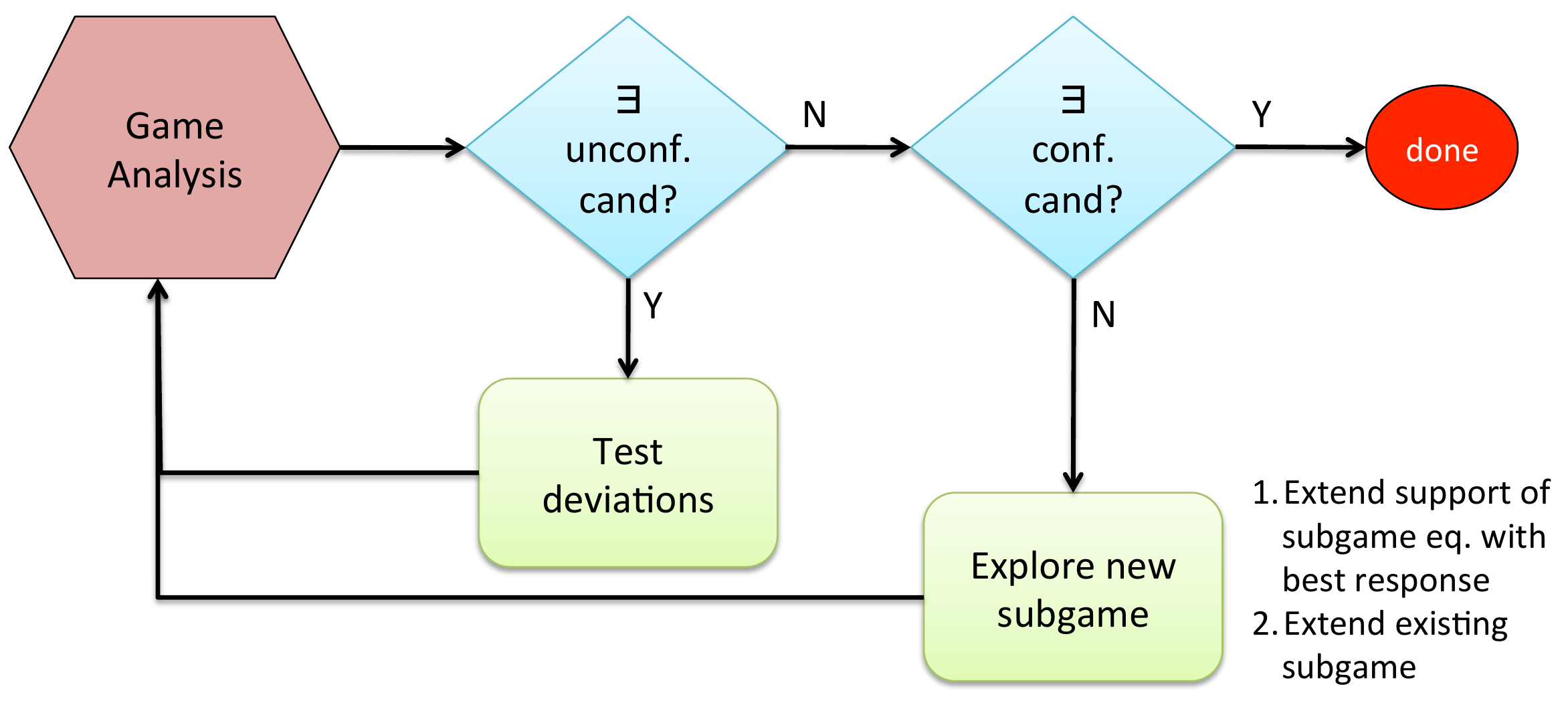}
 	\caption{EGTA inner loop.}
 	\label{fig:inner-loop}
\end{figure}

A high-level description of the EGTA inner loop is presented in Figure~\ref{fig:inner-loop}, with details provided in pseudocode as Algorithm~\ref{alg:inner-loop}. 
The loop operates with a fixed strategy space \( \stratcand=(\stratcand_1,\dotsc,\stratcand_R) \), specifying a set of available strategies for each role. 
The main input to the EGTA inner loop is \( G(\stratcand) \), the \emph{empirical game} comprised of payoff data accumulated thus far for profiles over \( \stratcand \)\@.
There are also several parametric inputs:
\begin{itemize}
	\item \rthresh, a regret threshold, 
	\item \minsamp, the minimum number of samples for initial evaluation, and  
	\item \mincsamp, the minimum number of samples for confirming evaluation.
\end{itemize}
On each iteration of the loop (as shown in the Figure), we run a game analysis algorithm over the current empirical game. 
Game analysis yields a set \( \solcand \) of candidate RSNE: role-symmetric profiles such that all payoffs have been evaluated (i.e., the empirical game has payoff estimates for all profiles in the support of the candidate), and there is no evaluated deviation with gain exceeding the regret threshold, \( \rthresh \).
We say that such a candidate is \emph{confirmed} if all possible deviations in the strategy space \( \stratcand \) have been evaluated.
The solution candidates can thus be partitioned into confirmed (\( \solcand_C \)) and unconfirmed (\( \solcand_U \)) candidates. 

To explain the third output of our game analysis procedure, it is necessary to describe its workings in more detail.
Standard game analysis (e.g., equilibrium-finding) algorithms require that the game form be fully specified: that  all payoffs be known.
Empirical games, however, generally have payoffs evaluated for only a subset of profiles.
To deal with this issue, our game analysis algorithm starts by identifying \emph{complete subgames}, that is, strategy subspaces \( \stratcand'=(\stratcand'_1,\dotsc,\stratcand'_R) \), with \( \stratcand'_r\subseteq \stratcand_r \), for all \( r \), such that all profiles over \( \stratcand' \) are evaluated in the empirical game.
We restrict attention to \emph{maximal} complete subgames, where adding any strategy to any role would render the subgame incomplete.
Note that if a profile \( q \) has regret bounded by \( \epsilon \) in game \( G(\stratcand) \), then it will also have regret at most \( \epsilon \) in any complete subgame \( G(\stratcand') \) for which \( \stratcand' \) contains the support of \( q \).
The converse is not true---a profile may have low regret in \( G(\stratcand') \) but be refuted by a deviation with respect to the larger strategy space \( \stratcand \).
Our game analysis thus simply runs standard equilibrium-finding methods for each maximal complete subgame, filters out the candidates \( \solcand_R \) that are refuted in the full profile space, and merges the rest to produce \( \solcand_C \) and \( \solcand_U \).

If the game analysis reveals unconfirmed candidates, we attempt to confirm or refute them by scheduling simulations of the unevaluated deviations. 
We specify the profiles to be scheduled in terms of the target profile set (i.e., support of the candidate), the deviation strategy sets (i.e., strategies outside that support, taken role-by-role), and the number of samples requested.  
Our EGTA infrastructure \citep{Cassell13mabs} manages a cumulative database of profile simulation results, and determines exactly how many new simulations of each relevant profile are required to reach the number requested.
Once collected, the new simulation results are added to the empirical game, which may of course change the candidate set. 
We therefore repeat the game analysis and deviation scheduling until there are no unconfirmed candidates.

If we have at least one confirmed candidate and no unconfirmed candidates, the inner loop terminates and returns \( \solcand_C \).
One fine point of our implementation of this procedure is that we require a greater number of samples, \( \mincsamp > \minsamp \), to label a candidate confirmed than we do to evaluate its candidate status initially.
Thus we ensure that all profiles in the support of purported confirmed solutions have at least \( \mincsamp \) samples, scheduling additional simulations if necessary and re-running game analysis to verify that the results still hold.

If there are no candidate solutions, then we need to further explore the current profile space. 
Given our game analysis algorithm, evaluating additional profiles cannot affect results unless the additional profiles lead to completion of a new maximal subgame.
Our first exploration effort aims to complete subgames that appear promising based on existing results.
Specifically, for any case where profile \( q \) is a \( \rthresh \)-RSNE of a maximal subgame \( G(\stratcand') \), but the best evaluated response to \( q \) is some strategy \( s_r\not\in\stratcand'_r \) in the larger game, we consider the subgame defined by adding \( s_r \) to the support of \( q \).
If all such subgames have already been explored yet still no confirmed equilibria have been found,%
\footnote{It may seem counterintuitive, but this can happen.
For example, consider a three-player symmetric game with three strategies \( \{A,B,C\} \) such that all profiles except \( (A,B,C) \) have been evaluated.
We could have a situation exhibiting a circular response pattern reminiscent of rock-paper-scissors, such as the following: \( (A,A,A) \) is the equilibrium of subgame \{A,B\}, with best-response \( C \); \( (B,B,B) \) is the equilibrium of subgame \{B,C\}, with best-response \( A \); and \( (C,C,C) \) is the equilibrium of subgame \{A,C\}, with best-response \( B \).
All best-response-enhanced subgames have been evaluated, but to find the true equilibrium (which has full support), we need the missing profile.
} 
the procedure then nondeterministically chooses to extend one of the current maximal subgames.
In the worst case, this process can lead to exhaustive exploration of the profile space, which necessarily contains at least one RSNE, which would be confirmed at that point.

To guarantee that the EGTA inner loop returns a confirmed solution candidate requires that the equilibrium-finding procedure is itself complete.  
Our implementation uses an incomplete method (replicator dynamics), so we may sometimes fail to find any RSNE through this process.

\input{inner}

\subsection{Strategy Exploration Algorithm (Outer Loop)}
\label{sec:outer}        

Completion of an EGTA inner loop leaves us in a state with one or more solutions that are confirmed with respect to the current strategy space, \( \stratcand \).
This space is a highly selective subset of the full strategy space, \strats, which is too large to consider wholesale in EGTA\@.
The function of \emph{strategy exploration} in compliance search (hexagons labeled ``Search for (non-)compliant deviation'' in Figure~\ref{fig:c-search}) is to augment \( \stratcand \) with promising strategies not yet included in the game-theoretic analysis.
One natural approach is to attempt to identify a beneficial deviation, or indeed the \emph{best-response} deviation in the broader strategy space \( \strats \) to a current solution candidate. 
For example, \citet{Phelps06} employed genetic algorithms to generate strategies for trading in continuous double auctions, evaluated by evolutionary criteria with respect to an existing strategy mix.
In an EGTA study in the same domain, \citet{Schvartzman09} employed reinforcement learning to search a large strategy space for an optimal response to the current equilibrium trading strategy. 
Best-response is generally not an ideal policy for introducing new strategies \citep{Jordan10sw}, as it completely prioritizes exploitation over exploration, and takes no account of performance with respect to strategies not currently in equilibrium.
Nevertheless, it has proven an effective heuristic in several domains, and the imperfection of optimization in practice does appear to provide a useful source of randomness in exploration.  

Our strategy exploration method likewise employs a heuristic best-response search, with the further refinement of directing the search alternately toward compliant or non-compliant deviations.
This explicit direction represents an exercise of \emph{due diligence}, ensuring that any conclusions based on solutions in one category (i.e., compliant or non-compliant) have been vetted by an express search for deviations in the complement category. 
The specific algorithm we employ for best-response search takes a stochastic hill-climbing approach, described in detail below. 

The description of our procedure assumes that strategies can be described as vectors of \( K \)   \emph{policy parameters}, \( \policy = \langle \pparam_1,\dotsc,\pparam_K \rangle\).%
\footnote{More general representations could be accommodated with modifications to the way that compliance is assessed, and how strategies are modified incrementally in local search.}   
For example, IBR policies include parameters for how much to increment or decrement reputations based on feedback reports, and reputation thresholds for deciding whether to accept an introduction.
Each parameter takes values in a specified domain, \( \pparam_i\in \pdomain_i \).
If all combinations are legal (also assumed here, for simplicity), then the cross-product of domains defines the overall strategy space, \( \strats = \pdomain_1 \times \dotsb \times \pdomain_K	 \). 
Let \( \compliance:\strats\to\Re \) be a function that evaluates the \emph{degree of compliance} of a strategy. 
Strategy \( \policy \) is considered \emph{compliant} if and only if \( \compliance(\policy) > 0	 \).
Note that using a graded compliance scale is optional; one could define a compliance function that simply maps all policies to \( \{-1,1\} \).

Our invocations of strategy exploration alternate between compliant and non-compliant modes.
For concreteness, we describe the process for \emph{non-compliant mode}, where the aim is to find a non-compliant deviation from a particular solution candidate.
The compliant mode simply reverses polarity in this description. 
Each run of strategy exploration focuses on a particular role \( \rho \), which we rotate through in successive invocations, taking care to switch compliance modes between consecutive runs for the same role.

The input solution profile for strategy exploration is selected from among the confirmed solutions \( X_C \) produced by the most recent EGTA inner loop.
We eliminate from consideration any \emph{closed} solutions, designated as such because they have already been fully explored.
In non-compliant mode, we select the most compliant open solution in \( X_C \). 
To evaluate compliance of a profile, we combine the compliance scores of its components.
Let \( \Pr(q_r=\policy) \) denote the probability that strategy \( \policy \) is played by players of role \( r \) in \( q \), and \( w_r \) a weight expressing the relative importance we accord to role~\( r \).
Then the compliance of profile \( q \) is given by
\[
 \compliance(q) = 
     \sum_{r=1}^R w_r \sum_{\policy\in\stratcand_r} \Pr(q_r=\policy)\compliance(\policy).
\]                                                
One natural role weighting (which we employ in the IBR study) is \( w_r=\numplayer{r} \), the number of players assigned to the role. 

The generation of new strategy candidates proceeds iteratively.
The heart of the algorithm is a procedure, \LocalSearch, that given a seed set of parametrically defined strategies, returns a set of incremental variations.
It aims to find variations that are close to one of the seed strategies, subject to the constraints that they be distinct from strategies already explored, and match the requested compliance status. 
The procedure must return at least one such variation, though the actual number may be variable or domain-dependent. 
The specific technique is also domain-dependent.
For IBR, \LocalSearch\ generates candidates by perturbing each of the parameters of each of the seed strategies by set increments, increasing this increment until at least one variation satisfies the constraints. 
The initial seed strategies are those supported in the candidate solution.

\begin{figure}[ht!]
  \centering
    \includegraphics[width=0.6\textwidth]{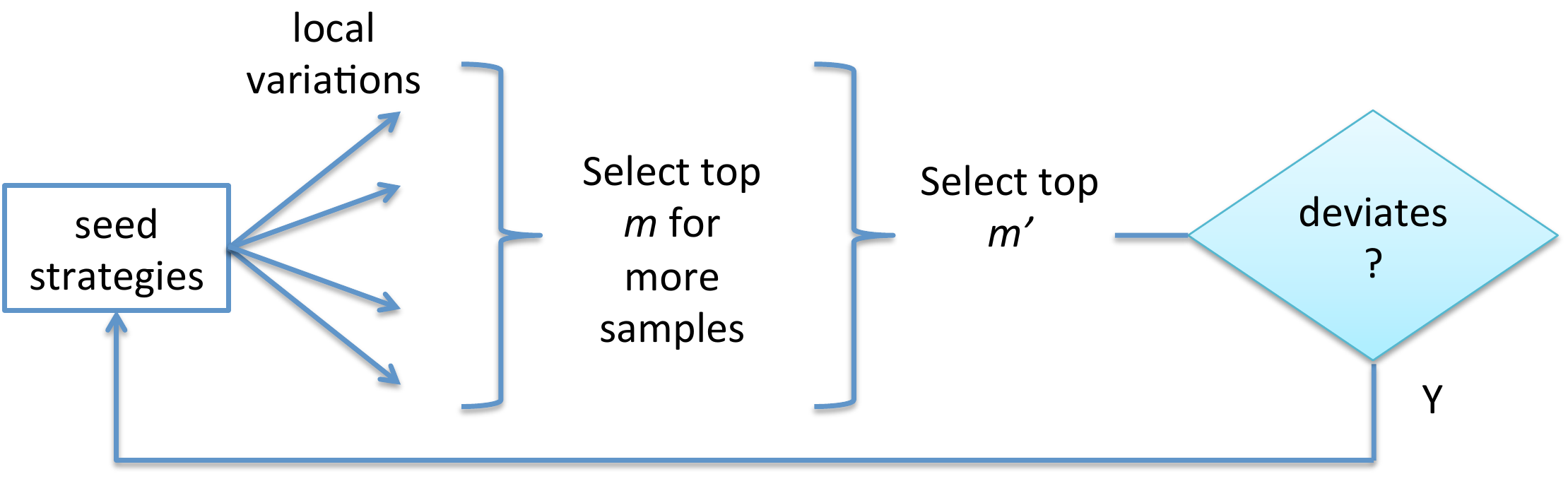}
 	\caption{Generating new strategy candidates by local search in parameter space.}
 	\label{fig:strategy-search}
\end{figure}

As illustrated in Figure~\ref{fig:strategy-search} and detailed in pseudocode by Algorithm~\ref{alg:outer-loop}, the generated variations are evaluated by scheduling deviation profiles for simulation.
Once these are complete, we select the \( m \) best of these for additional simulation samples.
Based on the refined deviation-gain estimates we obtain from this, we select the top \( m' \) of these as seed strategies for the next iteration.
For the IBR study, we take \( m=5 \) and \( m'=2 \).

In each iteration, we keep track of the highest payoff seen so far.
If this fails to improve, we terminate and return the best of the newly explored strategies.
If this is actually a beneficial deviation, it gets added to the candidate strategy set \( \stratcand_\rho \), and per the top-level search process (Figure~\ref{fig:c-search}), proceed back to the EGTA inner loop.
If it is not a beneficial deviation, we record failure and try again with a different role and compliance state.
If all roles and compliance states have been tried without success, then we mark this solution candidate as \emph{closed} and proceed to the next candidate.

\input{outer}

\section{IBR Compliance Analysis}
\label{sec:ibr-analysis}    

We evaluate our general compliance search procedure through a computationally intensive case study of introduction-based routing.
In this section, we describe how the general methods are specialized for IBR, and present the experimental setup and results of this study.

\subsection{Environments}
 
\begin{figure}[htb]
\begin{center}
\includegraphics[width=11cm]{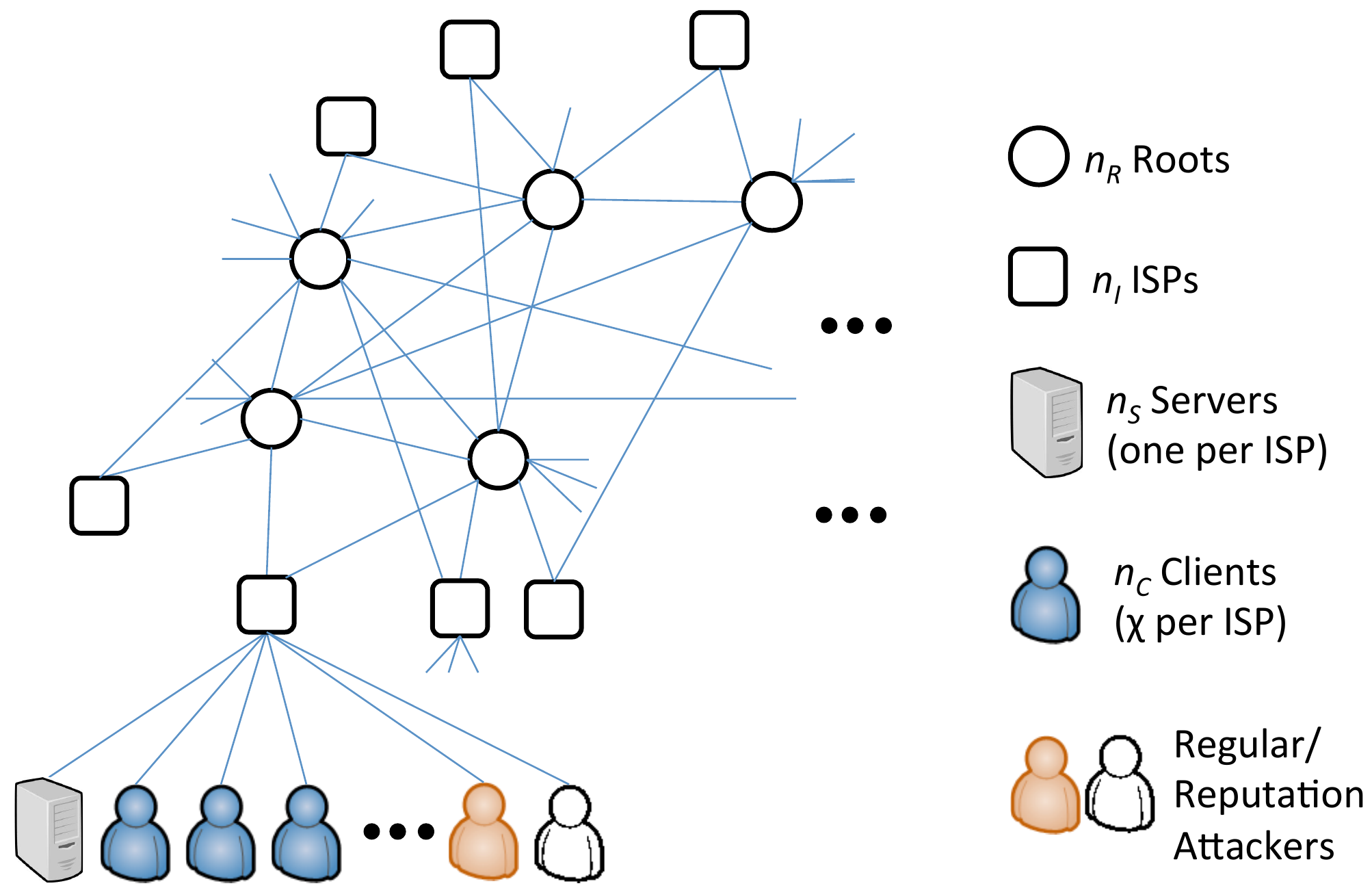}
\caption{The topology of a priori connections in the subject network.
}
\label{fig:network_topo}
\end{center}
\end{figure}

We conduct our analysis in three IBR game environments, sharing a basic network structure but differing on the numbers of nodes and players, and in attack model characteristics.
The network connection topology (Figure~\ref{fig:network_topo}) is a redundant tree, with \( \numnode{R} \) fully connected
introducers at the root and the clients and servers at the leaves.
Each client and server has a single a priori connection to an
introducer, notionally playing the role of ISP\@.
Thus, each client-server connection requires between one and four introductions to establish. 
There are \( \numnode{I} \) ISPs in the network, each connected to a randomly selected pair of the root introducers.
Each of the ISPs also has \( \cperisp \) clients and one server attached, for a total of \( \numnode{C} = \cperisp\numnode{I} \) client nodes and \( \numnode{S} = \numnode{I} \) server nodes.
The configuration parameters \( \numnode{R} \), \( \numnode{I} \), and \( \cperisp \) thus determine the numbers of nodes of every role, as well as the connectivity structure of the network.

A simulation run plays out an IBR scenario for 10,000 time units.
An active client generates a message, directed to a server selected uniformly at random, and then goes to sleep for a number of time units \( \sim U[0,39] \).
As a result, each client sends 330 messages in an average run.
In addition, \( \numnode{a} \) attacker nodes each select a random target at the start of simulation, after which messages sent to this target are attacks with probability 0.05.
Attacks are detected with probability 0.9, and succeed with probability 0.3 if not detected.
Non-attacks are falsely detected with probability 0.001.
A further set of \( \numnode{ra} \) reputation attackers generate false attack reports for a random target. 
Both forms of attacker are part of the network, but follow fixed behaviors and are thus not represented as players in the game.

Our experiments covered three network configurations, described in Table~\ref{tab:configs}.
For instance, the network configuration for environment~1 has \( \numnode{R}=7 \), \( \numnode{I}=49 \), and \( \cperisp=99 \), thus comprising 4956 nodes: seven roots, 49 ISPs, 49 servers, and 4851 clients.
A simulation of this configuration comprises an average of 3.2 million application messages and takes on the order of ten minutes on a single CPU per run.%
\footnote{Message volume and running times scale roughly proportionally with network size.} 
Beyond the network structure, all configurations employ the same simulation settings, except that in environment~1 the attackers choose only server targets, whereas in environments 2 and~3 they attack both clients and servers.

\begin{table}  
\centering
	\begin{tabular}{|c|cccccc|c|}
	\hline
	env & \( \numnode{R} (\numplayer{R}) \) & \( \numnode{I} (\numplayer{I}) \) & \( \cperisp (\numplayer{C}) \) & \( (\numplayer{S}) \) & \( \numnode{a} \) & \( \numnode{ra} \) & nodes (players)\\
	\hline
	1 & 7 (1) & 49 (1) & 99 (3) & (1) & 49 & 49 & 4956 (6)\\
	2 & 3 (1) & 18 (2) & 33 (3) & (2) & 48 & 16 & 633 (8)\\
	3 & 5 (1) & 30 (2) & 66 (2) & (1) & 90 & 30 & 2045 (6)\\
	\hline
	\end{tabular} 
	\caption{Network configurations employed in our IBR compliance analysis.
	The final column totals the non-attacking nodes---clients, servers, roots, and ISPs---and players in the game.}
\label{tab:configs}
\end{table}

Game-theoretic analysis with hundreds or thousands of players would be computationally infeasible, therefore we employ aggregation techniques to approximate the game by one with fewer players.
Specifically, we use \emph{hierarchical reduction} \citep{Wellman05rlcs}, which combines multiple nodes of a given role into single players.
With this reduction, for instance our 4956-agent environment is modeled as a six-player game, comprising:
\begin{itemize}
	\item three clients, each representing 1617 client nodes;
	\item one ISP introducer, representing the 49 ISP nodes; 
	\item one root introducer, representing the seven root nodes; and
	\item one server, representing the 49 server nodes.
\end{itemize}                                          
To simulate a profile in the reduced game, the strategy assigned to a given player is played by all nodes represented by that player.
The resulting payoff for the player is simply the average over payoffs to its constituent nodes.

\subsection{Simulation Coverage}

We perform our simulations on the computational cluster facility operated by the Center for Advanced Computing at the University of Michigan.
Our compliance search is implemented by a control script that employs our EGTAOnline facility \citep{Cassell13mabs}, a front-end that facilitates game-theoretic experiment management through an integration of cluster monitoring, database services, and analysis algorithms.
Between May 2012 and May 2013 we ran well over three million simulations of IBR network scenarios in service of this compliance analysis.             

Table~\ref{tab:simul} summarizes the extent of compliance search for our three IBR network environments.
For each environment, we report the number of iterations of the outer loop completed, and the scope of exploration.
Scope is measured by strategies in candidate set (sum over roles), the number of these that were auto-generated in the outer loop, profiles evaluated (number and percentage of profile space), and the total number of simulation runs.
These numbers do not include any simulation evaluation that was incurred in the outer loop for strategies that did not make it into the candidate set.

\begin{table}  
\centering
	\begin{tabular}{|c|cccccc|}
	\hline
	env & iters & strategies & auto-gen & profiles & \% profiles & sim runs \\
	\hline
	1 & 9 &   25 &  9  &  7808 & 36\% &  778k \\
	2 & 23 &  37 & 21  & 61272 &  2\% & 2566k \\
	3 & 33 &  29 & 21  &  2515 & 23\% &   54k \\
	\hline
	\end{tabular} 
	\caption{Compliance search over three environments.}
\label{tab:simul}
\end{table} 

The difference between columns ``strategies'' and ``auto-gen'' in Table~\ref{tab:simul} is the number of manually defined seed strategies included.  
In all environments, we included for each role a basic compliant policy supplied by the IBR protocol designers \citep{Frazier11}, and a non-compliant policy that makes introductions and accepts connections indiscriminately, oblivious to reputations and feedback. 
For environments 1 and~2, the initial sets include two additional compliant policies for each role.
The difference between columns ``iters'' and ``auto-gen'' in the table is the number of times the outer loop returned a strategy that was not a beneficial deviation from the target profile, and thus was not added to the strategy space.

The exploration and simulation process occurred over a period of many months, during which the procedure was not entirely constant.  
In particular, the ordering of various search steps, as well as some thresholds and other parameters were tuned in the course of running the compliance analysis of environment~1, and into the beginning part of our environment~2 analysis.  
Moreover, our empirical game for environment~1 includes many simulations that were manually scheduled in order to refine payoff estimates and accelerate the inner loop.
All runs of the outer loop were fully automated, as was the entire process for environments 2 and~3.  

Parameter settings (Table~\ref{tab:params}) were fixed for all of our analysis of environment~3, and most of the environment~2 analysis. 
Regret thresholds \( \rthresh \) represent a tiny fraction of payoffs, which generally exceed \( 10^5 \) for roots and ISPs, and \( 10^6 \) for clients and servers.
\begin{table}[h]
  \centering
  \begin{tabular}{|c|ccccccc|} \hline
    & & \multicolumn{2}{c}{inner loop} & \multicolumn{2}{c}{outer loop} & & \\
    env & $\rthresh$ & $\minsamp$ & $\mincsamp$ & $\minsamp$ & $\mincsamp$ & $m$ & $m'$ \\ \hline
    1 & 100 & 40 & 80 & 15 & 40 & 5 & 2 \\
    2 & 20 & 40 & 80 & 15 & 40 & 5 & 2 \\
    3 & 70 & 20 & 40 & 10 & 20 & 5 & 2 \\ \hline
  \end{tabular}
  \caption{Parameter settings employed in compliance search process.
\label{tab:params}}
\end{table}

\subsection{Results} 

We present detailed results for each of the IBR network environments studied, followed by an overall assessment.

\subsubsection{Environment 1 (6 players, 4956 nodes)}

The first environment studied is the largest network, almost 5000 nodes. 
The empirical game for this environment includes 7808 evaluated profiles, which cover 157 maximal subgames.                                                      
There are many confirmed solutions at the regret threshold \( \tau=100 \).
Table~\ref{tab:env1} presents, in increasing regret order, the 11 \( \epsilon \)-RSNE we identified%
\footnote{We include at most one solution (the best) with any particular support, and impose a minimum probability at which a strategy may be included in a mixed profile.
Any approximate RSNE is surrounded by a region of almost-as-good approximate equilibria in its neighborhood.}
with \( \epsilon\le 10 \).
We label by \agnt{C\( k \)} (\agnt{N\( k \)}) the compliant (non-compliant) strategy generated in iteration~\( k \) of the outer loop.
\agnt{N} (with no number) denotes the initial non-compliant (oblivious) strategy, \agnt{C} the initial designer-specified compliant strategy, and \agnt{C'} another initially supplied compliant strategy.
Mixed strategies are written \( [s_1,p_1;s_2,p_2;\dotsc] \), indicating strategy \( s_i \) is played with probability \( p_i \). 
The last probability is omitted as it is determined by the rest. 

\begin{table}[ht]
\begin{center}
\begin{tabular}{|c|c|c|c|c|}
\hline 
client & ISP & root & server & regret (\( \epsilon \)) \\ \hline
\( [\agnt{N3},.10; \agnt{N7},.34;\agnt{N}] \) &  \( [\agnt{N},.62;\agnt{C'}] \) & \agnt{N5} & \agnt{C6} & 0.0 \\
\( [\agnt{C},.33; \agnt{C2},.22;\agnt{N7}] \) & \agnt{N1} & \agnt{C} & \agnt{C6} & 0.0 \\
\( [\agnt{C},.16; \agnt{N3}] \) & \agnt{N1} & \( [\agnt{C},.68;\agnt{C8}] \) & \agnt{C6} & 0.2 \\
\( [\agnt{N3},.25; \agnt{N7},.43;\agnt{N}] \) &  \agnt{N} & \( [\agnt{C},.38;\agnt{C8},.10;\agnt{N5}] \) & \agnt{C6} & 1.0 \\
\( [\agnt{N3},.24; \agnt{N7},.41;\agnt{N}] \) &  \agnt{N} & \( [\agnt{C},.36;\agnt{N5}] \) & \agnt{C6} & 1.3 \\
\( [\agnt{C},.16; \agnt{C4},.11;\agnt{N3}] \) &  \agnt{C9} & \( [\agnt{C8},.48;\agnt{N5}] \) & \agnt{C6} & 5.8 \\
\( [\agnt{N3},.34; \agnt{N7}] \) &  \agnt{C'} & \agnt{C8} & \agnt{C6} & 6.7 \\
\( [\agnt{N3},.33; \agnt{N7},.66;\agnt{C'}] \) &  \agnt{C'} & \agnt{C8} & \agnt{C6} & 7.0 \\
\( [\agnt{C2},.32; \agnt{C'}] \) &  \( [\agnt{N},.74;\agnt{C'}] \) & \agnt{C8} & \agnt{C6} & 8.3 \\
\( [\agnt{C},.46; \agnt{N3}] \) & \agnt{C9} & \( [\agnt{C},.06;\agnt{C8},.18;\agnt{N5}] \) & \agnt{C6} & 8.6 \\
\( [\agnt{C4},.77; \agnt{N}] \) &  \agnt{C9} & \agnt{C} & \agnt{C6} & 9.8 \\
 \hline
\end{tabular}
\end{center}
\caption{Confirmed \( 10 \)-RSNE of environment~1, in ascending order of regret.
\label{tab:env1}}
\end{table} 

The first feature one might notice from Table~\ref{tab:env1} is that servers always play strategy \agnt{C6}.
Indeed, this is true for all approximate equilibria in all subgames that include \agnt{C6}.
This strategy was automatically generated by our exploration procedure (Algorithm~\ref{alg:outer-loop}), and is significantly better than all other server strategies evaluated, including the designer-supplied compliant strategy \agnt{C}.
Servers apparently have a strong incentive to comply with the IBR protocol, and an effective way to do so.
That servers exhibit the greatest compliance is perhaps not surprising in this environment, as they are the only nodes that are subject to attack. 

Compliance for other nodes is far less constant. 
Root nodes tend to comply in 8 of 11 of the RSNE, but clients and ISPs exhibit a fairly balanced choice between compliant and non-compliant strategies.
It appears that the indirect benefits from participating faithfully in IBR for nodes not subject to attack is fairly weak.
The costs, however, are also weak, as we do not see a very strong leaning toward non-compliance for any role.
Assessing overall compliance depends on the relative importance we accord the various roles.  
Clients represent 1/4 of the roles, but 3/6 of the players and 4851/4956 of the nodes.
Table~\ref{tab:env1-compliance} summarizes the overall fraction of compliance in the top equilibrium profiles, according to different ways of weighting the respective roles: evenly, by number of players, or number of nodes.

\begin{table}[ht]
\begin{center}
\begin{tabular}{|c|c|c|c|}
\hline     
soln & role & player & node \\ \hline  
1 & 34.5\% & 23.0\% &  1.4\% \\
2 & 63.7\% & 60.8\% & 54.9\% \\
3 & 79.0\% & 58.1\% & 17.9\% \\ 
4 & 37.0\% & 24.7\% &  1.1\% \\ 
5 & 33.9\% & 22.6\% &  1.0\% \\ 
6 & 68.7\% & 54.8\% & 28.5\% \\
7 & 75.0\% & 50.0\% &  2.1\% \\ 
8 & 75.2\% & 50.5\% &  3.1\% \\ 
9 & 81.6\% & 87.7\% & 99.3\%  \\ 
10 & 67.7\% & 60.6\% & 47.4\% \\
11 & 94.1\% & 88.3\% & 77.1\% \\ 
\hline
\end{tabular}
\end{center}
\caption{Overall fraction of compliance for each solution, for various role weightings.
\label{tab:env1-compliance}}
\end{table}          

\subsubsection{Environment 2 (8 players, 633 nodes)}  

Environment~2 incorporates three significant changes from environment~1.
First, the network is scaled down to about one-eighth the size.
Second, despite fewer nodes, we increased the number of players (i.e., adopted a less coarse hierarchical reduction), so that both ISPs and servers are represented by two players.
Third, we include attacks on clients as well as servers.
The numbers of attackers are scaled so that the level of attack on any eligible target is roughly the same.
(Reputation attacks are still addressed to servers only.) 

As indicated in Table~\ref{tab:simul}, the empirical game generated for this environment is considerably larger, covering over 60,000 profiles over 37 strategies (7--10 per role).
The analysis yields 13 10-RSNE, listed in Table~\ref{tab:env2}.
Names are as for environment~1, with the addition of \agnt{C''}, a designer-supplied compliant strategy constructed expressly to deal with reputation attacks.

\begin{table}[ht] 
	\begin{small}
\begin{center}
\begin{tabular}{|c|c|c|c|c|}
\hline 
client & ISP & root & server & \( \epsilon \) \\ \hline  
\agnt{C11} &  \( [\agnt{C4},.29;\agnt{C21},.10;\agnt{N1},.31;\agnt{N12}] \) & \( [\agnt{C'},.18;\agnt{C2},.31;\agnt{C9},.01;\agnt{N5},.31;\agnt{N16}] \) & \agnt{C''} & 4.7 \\
\agnt{C11} &  \( [\agnt{C15},.45;\agnt{N}] \) & \( [\agnt{C},.38;\agnt{C19}] \) & [\agnt{C''},.72;\agnt{C13}] & 5.2 \\
\agnt{C11} &  \( [\agnt{C'},.01;\agnt{C15},.45;\agnt{N}] \) & \( [\agnt{C},.38;\agnt{C19}] \) & [\agnt{C''},.72;\agnt{C13}] & 5.7 \\
\agnt{C11} &  \( [\agnt{C4},.29;\agnt{C21},.07;\agnt{N1},.27;\agnt{N12}] \) & \( [\agnt{C},.21;\agnt{C2},.56;\agnt{C9},.07;\agnt{N}] \) & \agnt{C''} & 7.0 \\
\agnt{C11} &  \( [\agnt{C4},.29;\agnt{C21},.07;\agnt{N1},.27;\agnt{N12}] \) & \( [\agnt{C'},.05;\agnt{C2},.31;\agnt{N5},.62;\agnt{N16}] \) & \agnt{C''} & 7.1 \\
\agnt{C11} &  \( [\agnt{C4},.24;\agnt{C21},.03;\agnt{N1},.35;\agnt{N12}] \) & \( [\agnt{C2},.38;\agnt{N5}] \) & \agnt{C''} & 7.5 \\
\agnt{C11} &  \( [\agnt{C'},.02;\agnt{C4},.72;\agnt{N12}] \) & \( [\agnt{C},.30;\agnt{C9},.31;\agnt{N16}] \) & \agnt{C13} & 7.9 \\
\agnt{C11} &  \( [\agnt{C'},.31;\agnt{N},.45;\agnt{N12}] \) & \( [\agnt{C2},.30;\agnt{N16}] \) & [\agnt{C''},.25;\agnt{C6}] & 8.3 \\
\agnt{C11} &  \( [\agnt{C'},.10;\agnt{C4},.57;\agnt{N},.17;\agnt{N12}] \) & \multicolumn{1}{|l|}{\( [\agnt{C},.17;\agnt{C''},.22;\agnt{C2},.25;\agnt{N5},.28; \)} & \agnt{C6} &  \\  
& &  \multicolumn{1}{|r|}{\( \agnt{N16},.06;\agnt{N22}] \)} & & 9.1 \\
\multicolumn{1}{|l|}{\( [\agnt{C11},.80; \)} &  \( [\agnt{C'},.29;\agnt{C4},.65;\agnt{N12}] \) & \( \agnt{C2} \) & \agnt{C6} &  \\
\multicolumn{1}{|r|}{\( \agnt{C17}] \)} & &  & & 9.3 \\
\agnt{C11} &  \( [\agnt{C'},.08;\agnt{C4},.55;\agnt{N},.16;\agnt{N12}] \) & \( [\agnt{C},.04;\agnt{C''},.18;\agnt{C2},.37;\agnt{N5},.35;\agnt{N16}] \) & \agnt{C6} & 9.7 \\
\agnt{C11} &  \( [\agnt{C'},.36;\agnt{N},.01;\agnt{N1},.25;\agnt{N12}] \) & \( [\agnt{C},.25;\agnt{C'},.01;\agnt{C2}] \) & \agnt{C''} & 9.9 \\
\agnt{C11} &  \( [\agnt{C12},.32;\agnt{N}] \) & \( [\agnt{C},.23;\agnt{N16},.65;\agnt{N22}] \) & [\agnt{C''},.23;\agnt{C6}] & 9.9 \\
 \hline
\end{tabular}
\end{center} 
\end{small}
\caption{Confirmed \( 10 \)-RSNE of environment~2, in ascending order of regret.
\label{tab:env2}}
\end{table} 

Once again, we see that the nodes that are subject to attack---clients as well as servers in this case---are universally compliant, and relatively consistent in strategy across the top confirmed solutions.  
The intermediaries who serve only as introducers in these scenarios have a much more varied strategy selection.
The overall compliance fractions are presented in Table~\ref{tab:env2-compliance}.

\begin{table}[ht]
\begin{center}
\begin{tabular}{|c|c|c|c|}
\hline     
soln & role & player & node \\ \hline  
1 & 72.2\% & 78.5\% & 98.0\% \\   
2 & 86.4\% & 86.4\% & 98.4\% \\                         
3 & 86.4\% & 86.4\% & 98.5\% \\ 
4 & 72.7\% & 74.6\% & 97.3\% \\ 
5 & 68.2\% & 76.2\% & 97.9\% \\   
6 & 66.1\% & 73.9\% & 97.6\% \\ 
7 & 83.8\% & 88.6\% & 99.1\% \\  
8 & 65.3\% & 74.0\% & 97.7\% \\  
9 & 82.7\% & 87.1\% & 98.9\% \\  
10 & 98.3\% & 98.3\% & 99.8\% \\  
11 & 80.7\% & 85.7\% & 98.8\% \\  
12 & 84.1\% & 84.1\% & 98.2\% \\  
13 & 55.8\% & 65.4\% & 96.8\% \\  
\hline
\end{tabular}
\end{center}
\caption{Overall fraction of compliance for each solution, for various role weightings.
\label{tab:env2-compliance}}
\end{table}          

\subsubsection{Environment 3 (6 players, 2045 nodes)} 

The third environment is intermediate in size between the other two.
It is defined for six players: two clients, two ISPs, one root, and one server.
As for environment~2, both clients and servers are subject to attack.

This environment was evaluated through 33 iterations, albeit with many fewer simulation samples than the previous environments.
The result is a single confirmed equilibrium at \( \rthresh=70 \).
The solution is a PSNE, that is a pure-strategy profile with zero regret:
\begin{center}
\begin{tabular}{|c|c|c|c|c|}
\hline 
client & ISP & root & server & regret (\( \epsilon \)) \\ \hline 
\agnt{C33} & \agnt{N} & \agnt{N6} & \agnt{N4} & 0.0 \\
\hline
\end{tabular}
\end{center}                                         

In contrast to the other environments, here the servers are non-compliant, even though subject to attack.
The overall compliance for this profile is 25.0\% by role, 33.3\% by player, and 96.8\% by node.
Although we lack a definitive explanation for the non-compliance of servers, there are several reasons to view this result as providing only weak evidence against the general compliance of nodes under attack.
First, on examination of its specific policy parameters, we find server strategy \agnt{N4} is actually close to the borderline we defined between compliant and non-compliant strategies.
In fact it is partially compliant, as it does adjust reputations to some degree based on reports, and does regard these reputations in decision making.  
It is classified as non-compliant because these adjustments do not meet the levels we had (somewhat arbitrarily) defined as required for compliance.
Second, strategy \agnt{N4} is only slightly preferred to compliant server strategy \agnt{C23} (\( \epsilon = 22 \)) given the other players' strategies in this profile.
In contrast, the best non-compliant client deviation for this profile has regret of 1275.
Finally, the result seems sensitive to the client strategy \agnt{C33}, which was introduced in the final iteration of this evaluation.
If we omit this strategy, the sole equilibrium is the following PSNE, in which all roles but the ISP are compliant:
\begin{center}
\begin{tabular}{|c|c|c|c|c|}
\hline 
client & ISP & root & server & regret (\( \epsilon \)) \\ \hline 
\agnt{C27} & \agnt{N} & \agnt{C9} & \agnt{C29} & 0.0 \\
\hline
\end{tabular}
\end{center}                                         

%

\subsubsection{Discussion}

We find across all three environments an incentive for substantial but not universal compliance.
Given that following the protocol incurs a modest cost in forgoing connections that might prove valuable, the compliance we see is evidence for its benefits in attack resistance.
Incentives to comply tend to be quite strong for nodes subject to attack, and relatively weak for intermediary nodes.                                                                      
That full compliance is never an equilibrium is not very surprising. 
Intuitively, once a critical fraction of other agents are complying with IBR, the network is protected enough that remaining agents need not comply in order to preserve network performance.  
Another interpretation is that the marginal security benefit of compliance is diminishing beyond a certain point.
Further extension and analysis of the empirical game may yield a more precise characterization of this phenomenon, and the general shape of costs and benefits of IBR compliance.

\section{Conclusion} 

The goal of this effort was to develop a systematic way to evaluate incentive properties of complex network protocols, employing large-scale simulation in service of strategic analysis according to game-theoretic principles.
Our hypothesis is that we could organize a process of searching over a large set of policy profiles by focusing attention on regions that are most relevant to resolving the compliance/non-compliance distinction.  
The procedure we defined and implemented addresses two key search issues: control of simulation over a fixed strategy space (inner loop), and exploration of a broader policy space to identify new promising strategies (outer loop).                                                           

As a case study, we applied this procedure to evaluate compliance with an interesting network security protocol: introduction-based routing.
The IBR study drove refinement of the algorithm, and enabled us to exercise both of its key components.
Though far from exhaustive, this investigation provided clear evidence that participants vulnerable to attack have substantial incentive to comply with the protocol, and this in conjunction with partial compliance by others is sufficient to achieve the protocol's security aims.
The results suggest that further attention should be paid to enhancing incentives to intermediary nodes, and the simulation data provides a basis for a more in-depth game-theoretic analysis.

Further research is required to understand the procedure's suitability for other domains, and to assess its strengths and weaknesses. 
One natural question is robustness: whether in fact qualitative conclusions about compliance are insensitive to specific choices in representation, search control, and search extent.
Preliminary evidence suggests that the qualitative results are indeed relatively stable, particularly as compared to game-theoretic conclusions about specific policy instances.
Another issue to address is the computationally intensiveness of the procedure.
On the one hand, part of the whole idea is to take advantage of large-scale computational resources that are increasingly available for efforts of this kind.
On the other, we would always prefer to get the most power from any given amount of simulation effort, and no doubt further experience will be required to identify the best ways to direct that effort toward the greatest impact on strategic insight.

\subsection*{Acknowledgments}

The IBR policies and simulation facility were developed in collaboration with Gregory Frazier and Edward Petersen. 
We benefited from many discussions about IBR with these individuals as well as O.~Patrick Kreidl. 
Ben-Alexander Cassell and Bryce Wiedenbeck extended their simulation management and game analysis software, respectively, to facilitate this project.
Thanks also to Erik Brinkman and anonymous reviewers for helpful comments on earlier drafts.

\bibliographystyle{plainnat}
\begin{small}
	\bibliography{ibrcomp}
\end{small}

\end{document}

%% file: inner.tex
\begin{algorithm}
\caption{EGTA Inner Loop}
\label{alg:inner-loop}
\begin{algorithmic}
\Require Strategy space $\stratcand=(\stratcand_1,\dotsc,\stratcand_R)$, Empirical game $G(\stratcand)$, Regret threshold $\rthresh$, Numbers of required samples \( \minsamp \) and \( \mincsamp \)
\Ensure Confirmed mixed RSNE \( \{q: \epsilon(q) \le \rthresh \mathrm{; all\ deviations\ of\ } q \mathrm{\ evaluated}\}  \)

\State
\State $\mathit{foundConfirmed} \gets \mathbf{false}$ 
\While{ $\neg\mathit{foundConfirmed}$} 
	\Repeat
		\State $(\solcand_C,\solcand_U,\solcand_R) \gets$ $\mathit{GameAnalysis}(G(\stratcand), \rthresh)$
	\For{$q \in \solcand_U$}
		\State $\mathit{ScheduleDeviation}(\Support(q), \stratcand\setminus \mathit{StrategiesIn}(q),\minsamp)$
	\EndFor
	\Until{$\solcand_U$ is empty}
	\State	
	\If{$\solcand_C \neq \emptyset$}
		\State $\mathit{foundConfirmed} \gets \mathbf{true}$
		\For {$q \in \solcand_C$}
			\If {$\mathit{foundConfirmed}\wedge \mathit{minSamples}(\Support(q)) < \mincsamp$}
				\State $\mathit{ScheduleSubgame}(\mathit{StrategiesIn}, \mincsamp)$ 
				\State $\mathit{foundConfirmed} \gets \mathbf{false}$
 			\EndIf
		\EndFor

    \Else
        \State \( \mathit{explored}\gets \mathbf{false} \) 
 	    \For {$q \in \solcand_R$} 
	       \If {\( \neg\mathit{explored}\wedge \neg\mathit{Subgame}(\mathit{StrategiesIn}(q) + \mathit{BestResponse}(q)) \)}  
	           \State \( \mathit{explored}\gets \mathbf{true} \)
	           \State $\mathit{ScheduleSubgame}(\mathit{StrategiesIn}(q) + \mathit{BestResponse}(q), \mincsamp)$
           \EndIf
	     \EndFor
     \If {\( \neg\mathit{explored} \)}        
		\State $\mathit{ScheduleSubgame}(\mathit{incrSubgame}(), \minsamp)$
	\EndIf  
	\EndIf
\EndWhile 
\State \Return \( \solcand_C \)
\end{algorithmic}
\end{algorithm} 

Notes to Algorithm~\ref{alg:inner-loop}:  
\begin{enumerate}
	\item $\Support(q)$ denotes the support of a profile $q$.
	\item $\mathit{ScheduleDeviation}(Q, \mathcal{S},n)$ generates simulations for profiles formed by deviating from profiles in set \( Q \) to strategies in role:strategy mapping \( \mathcal{S} \), until at least \( n \) of each are accumulated.     
	\item $\mathit{ScheduleSubgame}(\mathcal{S}, n)$ generates simulations for profiles comprising strategies in role:strategy mapping \( \mathcal{S} \), until at least \( n \) of each are accumulated.     
\item $\mathit{incrSubgame}()$ returns a minimal strategy set that is not a complete subgame in the current empirical game.       
\item At termination, the empirical game over \( \stratcand \) is augmented such that all RSNE candidates are confirmed or refuted.
\end{enumerate}

%% file: outer.tex
\begin{algorithm}
\caption{Strategy Exploration Outer Loop (non-compliant mode, role~\( \rho \))}
\label{alg:outer-loop}
\begin{algorithmic}
\Require Empirical game $G(\stratcand)$, 
Open confirmed solution candidates \( \solcand_C \), 
Numbers of required samples \( \minsamp \) and \( \mincsamp \),
Numbers of strategies to select \( m \) and \( m' \)
\Ensure New compliant strategy to add to $\stratcand_\rho$
\State
\State $q \gets \arg\max_{q\in \solcand_C}\compliance(q)$    
\State \( \mathit{SeedStrategies}\gets \Support(q_\rho) \)  
\State \( \mathit{Explored}\gets \stratcand_\rho \)  
\State \( \mathit{newPayoff}\gets -\infty \)  
                                                          
\Repeat
  \State \( \mathit{bestPayoff}\gets \mathit{newPayoff} \)  
  \State \( \mathit{CandSet}\gets \LocalSearch(\mathit{SeedStrategies},\mathit{Explored},\mathbf{non\-compliant}) \)
  \State $\mathit{ScheduleDeviation}(\Support(q), \mathit{CandSet},\minsamp)$
  \State \( \mathit{Explored}\gets \mathit{Explored}\cup\mathit{CandSet} \) 
  \State $\mathit{ScheduleDeviation}(\Support(q), \mathit{SelectBestDeviators}(q,\rho,\mathit{CandSet},m), \mincsamp)$
  \State  \( \mathit{SeedStrategies}\gets \mathit{SelectBestDeviators}(q,\rho,\mathit{CandSet},m') \)
  \State \( \mathit{newPayoff}\gets \max_{s\in\mathit{SeedStrategies}}(u_\rho(s,q)) \)
\Until {\( \mathit{newPayoff} \le \mathit{bestPayoff} \)}
\State \Return \( \mathit{SelectBestDeviators}(q, \rho, \mathit{Explored}\setminus\stratcand_\rho, 1) \)		
\end{algorithmic} 
\end{algorithm}

Notes to Algorithm~\ref{alg:outer-loop}:
\begin{enumerate}  
	\item \( \LocalSearch(S_1,S_2,c) \) returns a nonempty set of strategy variations close to some strategy in \( S_1 \), distinct from \( S_2 \), and consistent with compliance state \( c \).
	\item \( u_\rho(s,q) \) denotes the payoff to a player of role \( \rho \) playing strategy \( s \) when the remaining players play according to profile \( q \).
	\item \( \mathit{SelectBestDeviators}(q,\rho,CS,m) \) evaluates the payoff to a role~\( \rho \) player deviating from profile \( q \) by playing \( s \), for each strategy \( s\in CS \). 
	It then returns the \( m \) best strategies according to this evaluation, or the entire set \( CS \) if \( \abs{CS}\le m \).
\end{enumerate}

%% file: ibrcomp.bbl
\begin{thebibliography}{23}
\providecommand{\natexlab}[1]{#1}
\providecommand{\url}[1]{\texttt{#1}}
\expandafter\ifx\csname urlstyle\endcsname\relax
  \providecommand{\doi}[1]{doi: #1}\else
  \providecommand{\doi}{doi: \begingroup \urlstyle{rm}\Url}\fi

\bibitem[Al-Bayaty and Kreidl(2013)]{AlK2013}
Richard Al-Bayaty and O.~Patrick Kreidl.
\newblock On optimal decisions in an introduction-based reputation protocol.
\newblock In \emph{38th International Conference on Acoustics, Speech, and
  Signal Processing}, 2013.

\bibitem[{\v C}agalj et~al.(2005){\v C}agalj, Ganeriwal, Aad, and
  Hubaux]{Cagalj05}
Mario {\v C}agalj, Saurabh Ganeriwal, Imad Aad, and Jean-Pierre Hubaux.
\newblock On selfish behavior in {CSMA/CA} networks.
\newblock In \emph{24th IEEE International Conference on Computer
  Communications}, pages 2513--2524, 2005.

\bibitem[Camerer(2003)]{Camerer03}
Colin~F. Camerer.
\newblock \emph{Behavioral Game Theory: Experiments in Strategic Interaction}.
\newblock Princeton University Press, 2003.

\bibitem[Cassell and Wellman(2013)]{Cassell13mabs}
Ben-Alexander Cassell and Michael~P. Wellman.
\newblock {EGTAOnline}: An experiment manager for simulation-based game
  studies.
\newblock In \emph{Multi-Agent Based Simulation XIII}, volume 7838 of
  \emph{Lecture Notes in Artificial Intelligence}. Springer, 2013.

\bibitem[Chen and Leneutre(2007)]{Chen07l}
Lin Chen and Jean Leneutre.
\newblock Selfishness, not always a nightmare: Modeling selfish {MAC} behaviors
  in wireless mobile ad hoc networks.
\newblock In \emph{27th International Conference on Distributed Computing
  Systems}, 2007.

\bibitem[Fearnley et~al.(2013)Fearnley, Gairing, Goldberg, and
  Savani]{Fearnley13}
John Fearnley, Martin Gairing, Paul Goldberg, and Rahul Savani.
\newblock Learning equilibria of games via payoff queries.
\newblock In \emph{Fourteenth ACM Conference on Electronic Commerce}, 2013.

\bibitem[Frazier et~al.(2011)Frazier, Duong, Wellman, and Petersen]{Frazier11}
Gregory Frazier, Quang Duong, Michael~P. Wellman, and Edward Petersen.
\newblock Incentivizing responsible networking via introduction-based routing.
\newblock In \emph{Fourth International Conference on Trust and Trustworthy
  Computing}, pages 279--293, 2011.

\bibitem[Grossklags et~al.(2008)Grossklags, Christin, and Chuang]{grossklags08}
Jens Grossklags, Nicolas Christin, and John Chuang.
\newblock Secure or insure?: A game-theoretic analysis of information security
  games.
\newblock In \emph{Seventeenth International Conference on World Wide Web},
  pages 209--218, 2008.

\bibitem[Han et~al.(2011)Han, Niyato, Saad, Ba\c{s}ar, and
  Hj{\o}rungnes]{Han11}
Zhu Han, Dusit Niyato, Walid Saad, Tamer Ba\c{s}ar, and Are Hj{\o}rungnes.
\newblock \emph{Game Theory in Wireless and Communication Networks: Theory,
  Models, and Applications}.
\newblock Cambridge University Press, 2011.

\bibitem[Jordan et~al.(2008)Jordan, Vorobeychik, and Wellman]{Jordan08vw}
Patrick~R. Jordan, Yevgeniy Vorobeychik, and Michael~P. Wellman.
\newblock Searching for approximate equilibria in empirical games.
\newblock In \emph{Seventh International Conference on Autonomous Agents and
  Multi-Agent Systems}, pages 1063--1070, 2008.

\bibitem[Jordan et~al.(2010)Jordan, Schvartzman, and Wellman]{Jordan10sw}
Patrick~R. Jordan, L.~Julian Schvartzman, and Michael~P. Wellman.
\newblock Strategy exploration in empirical games.
\newblock In \emph{Ninth International Conference on Autonomous Agents and
  Multi-Agent Systems}, pages 1131--1138, 2010.

\bibitem[Leyton-Brown and Shoham(2008)]{Leyton-Brown08}
Kevin Leyton-Brown and Yoav Shoham.
\newblock \emph{Essentials of Game Theory: A Concise Multidisciplinary
  Introduction}.
\newblock Morgan and Claypool, 2008.

\bibitem[Phelps et~al.(2006)Phelps, Marcinkiewicz, Parsons, and
  McBurney]{Phelps06}
S.~Phelps, M.~Marcinkiewicz, S.~Parsons, and P.~McBurney.
\newblock A novel method for automatic strategy acquisition in $n$-player
  non-zero-sum games.
\newblock In \emph{Fifth International Joint Conference on Autonomous Agents
  and Multi-Agent Systems}, pages 705--712, 2006.

\bibitem[Phelps et~al.(2010)Phelps, McBurney, and Parsons]{Phelps10}
Steve Phelps, Peter McBurney, and Simon Parsons.
\newblock Evolutionary mechanism design: A review.
\newblock \emph{Autonomous Agents and Multi-Agent Systems}, 21:\penalty0
  237--264, 2010.

\bibitem[Roy et~al.(2010)Roy, Ellis, Shiva, Dasgupta, Shandilya, and Wu]{Roy10}
Sankardas Roy, Charles Ellis, Sajjan~G. Shiva, Dipankar Dasgupta, Vivek
  Shandilya, and Qishi Wu.
\newblock A survey of game theory as applied to network security.
\newblock In \emph{43rd Hawaii International Conference on System Sciences},
  2010.

\bibitem[Schvartzman and Wellman(2009)]{Schvartzman09}
L.~Julian Schvartzman and Michael~P. Wellman.
\newblock Stronger {CDA} strategies through empirical game-theoretic analysis
  and reinforcement learning.
\newblock In \emph{Eighth International Conference on Autonomous Agents and
  Multi-Agent Systems}, pages 249--256, 2009.

\bibitem[Srivastava et~al.(2006)Srivastava, Neel, MacKenzie, Menon, DaSilva,
  Hicks, Reed, and Gilles]{Srivastava05}
Vivek Srivastava, James~A. Neel, Allen~B. MacKenzie, Rekha Menon, Luiz~A.
  DaSilva, James~E. Hicks, Jeffrey~H. Reed, and Robert~P. Gilles.
\newblock Using game theory to analyze wireless ad hoc networks.
\newblock \emph{IEEE Communications Surveys \& Tutorials}, 7\penalty0
  (4):\penalty0 46--56, 2006.

\bibitem[Sureka and Wurman(2005)]{Sureka:2005uq}
Ashish Sureka and Peter~R. Wurman.
\newblock Using tabu best-response search to find pure strategy {Nash}
  equilibria in normal form games.
\newblock In \emph{Fourth International Joint Conference on Autonomous Agents
  and Multi-Agent Systems}, pages 1023--1029, 2005.

\bibitem[Vratonjic et~al.(2010)Vratonjic, Hubaux, Raya, and
  Parkes]{Vratonjic10}
Nevena Vratonjic, Jean-Pierre Hubaux, Maxim Raya, and David~C. Parkes.
\newblock Security games in online advertising: Can ads help secure the web?
\newblock In \emph{Ninth Workshop on the Economics of Information Security},
  2010.

\bibitem[Vratonjic et~al.(2012)Vratonjic, Manshaei, Grossklags, and
  Hubaux]{Vratonjic12}
Nevena Vratonjic, Mohammad~Hossein Manshaei, Jens Grossklags, and Jean-Pierre
  Hubaux.
\newblock Ad-blocking games: Monetizing online content under the threat of ad
  avoidance.
\newblock In \emph{Eleventh Workshop on the Economics of Information Security},
  2012.

\bibitem[Walsh et~al.(2003)Walsh, Parkes, and Das]{Walsh03}
William~E. Walsh, David Parkes, and Rajarshi Das.
\newblock Choosing samples to compute heuristic-strategy {Nash} equilibrium.
\newblock In \emph{AAMAS-03 Workshop on Agent-Mediated Electronic Commerce},
  2003.

\bibitem[Wellman(2006)]{Wellman06}
Michael~P. Wellman.
\newblock Methods for empirical game-theoretic analysis (extended abstract).
\newblock In \emph{Twenty-First National Conference on Artificial
  Intelligence}, pages 1552--1555, 2006.

\bibitem[Wellman et~al.(2005)Wellman, Reeves, Lochner, Cheng, and
  Suri]{Wellman05rlcs}
Michael~P. Wellman, Daniel~M. Reeves, Kevin~M. Lochner, Shih-Fen Cheng, and
  Rahul Suri.
\newblock Approximate strategic reasoning through hierarchical reduction of
  large symmetric games.
\newblock In \emph{Twentieth National Conference on Artificial Intelligence},
  pages 502--508, 2005.

\end{thebibliography}
